\documentstyle[osa,manuscript]{revtex}

\begin{document}

\title
{Optical second harmonic generation probe of two-dimensional
ferroelectricity}

\author {O.A. Aktsipetrov\footnote{Corresponding author. O.A. Aktsipetrov's e-mail address
is aktsip@astral.ilc.msu.su. URL:  http://kali.ilc.msu.su}, T.V. Misuryaev, T.V. Murzina}
\address{Department of Physics, Moscow State University, Moscow 119899,
Russia}
\author {L.M. Blinov, V.M. Fridkin, S.P. Palto}
\address{Institute of Crystallography, Russian Academy of Sciences, Moscow 117333,
Russia}

\date{\today}
\maketitle

\begin{abstract}
{Optical second harmonic generation (SHG) is used as a noninvasive probe of two-dimensional (2D)
ferroelectricity in Langmuir-Blodgett (LB) films of copolymer vinylidene fluoride with
trifluorethylene. The surface 2D ferroelectric-paraelectric phase transition in the topmost layer of LB
films and a thickness independent (almost 2D) transition in the bulk of these films are observed in
temperature studies of SHG.}
\end{abstract}

\vspace{5mm} \noindent

\newpage
The role of dimensionality in phase transitions and ferroelectricity and ferromagnetism has been
one of the central points of basic theoretical and experimental studies for decades \cite{c1}. Significant
progress in the experimental studies of two-dimensional (2D) magnetic systems has been achieved
with molecular beam epitaxy (MBE) fabrication of magnetic monolayers \cite{c2}. In contrast to 2D
magnetic materials, serious technological problems still remain in the fabrication of ultrathin
ferroelectric films, even in the case of MBE. It has appeared recently that the Langmuir-Blodgett (LB)
technique is more successful than MBE in preparation of ultrathin ferroelectric films \cite{c3}. The LB
method was successfully used for preparation of noncentrosymmetric and polar films with perspective
pyroelectric, electro-optical \cite{c4}, and nonlinear optical properties \cite{c5}. Moreover, in the early 1970's
magnetic monolayers were fabricated by LB techniques \cite{c6}.

The first LB films based on poly-vinylidene fluoride (PVDF) have been prepared in the very late
1990's and recent results show that ultrathin LB films of this polymer behave as 2D ferroelectrics
\cite{c3,c7}. The LB technique has opened a new experimental chapter in 2D ferroelectricity which has
previously only been discussed theoretically. As for experimental studies, up to now ferroelectric
properties of these 2D structures were investigated by traditional measurements of dielectric
parameters. These methods usually entail the deposition of a cap electrode on top of a delicate layered
structure consisting of only a few polymer monolayers. As a result of this, a question always arises
how non-invasive such a deposition is and how undisturbed and reliable these measurements are.

Optical second harmonic generation (SHG) has been shown to be a rather simple, non-invasive
and informative probe to study the structure, symmetry and morphology of interfaces and ultrathin
(down to monolayers) films \cite{c8}. This advantage of SHG stems from its unique sensitivity to the
breakdown of crystallographic symmetry. In turn, this high sensitivity of SHG arises from the
symmetry selection rules for the second-order susceptibility: in the dipole approximation, SHG is
strongly forbidden in the bulk of centrosymmetric materials \cite{c8}, being allowed for structures with a
lack of inversion symmetry. For example, the application of external DC-electric field breaks down the
inversion symmetry of centrosymmetric materials. As a consequence, DC-electric field induced SHG is
allowed \cite{c9}. By analogy with external DC-electric field, a spontaneous polarization (SP) in
ferroelectrics breaks down the inversion symmetry and produces strong dipole SHG. Thus, on the one
hand, SHG is a very sensitive probe of ferroelectricity and ferroelectric-paraelectric phase transitions,
as the symmetry of a material changes from a centrosymmetric to a noncentrosymmetric type as
spontaneous polarization appears. On the other hand, this is an electrode-free method which is non-
invasive for ultrathin films.

In this paper, the ferroelectric properties and ferroelectric-paraelectric phase transition are
studied in ultrathin electrode-free PVDF LB films by means of a temperature-dependent SP induced
SHG.

Ferroelectric LB films composed from the copolymer vinylidene fluoride with trifluorethylene
(P(VDF-TrFE)(70:30 mol \%)) were deposited onto a fused quartz substrate. The thickness of a single
monolayer of P(VDF-TrFE) LB film is approximately 0.5 nm. The 15-monolayer-thick and 60-
monolayer-thick LB films were studied. For the SHG measurements the output of a Q-switched
YAG:Nd$^{3+}$ laser at a wavelength of 1064-nm was used with a pulse duration of 15 ns, and an intensity
of 1 MW/cm$^2$ focused onto a spot approximately 0.5 mm in diameter at an angle of incidence of 45$^o$ .
The reflected SHG signal is selected by a monochromator and detected by a PMT with an angular
aperture of the detection system of approximately 10$^{-1}$ sr and gated electronics. The s-in, s-out and p-in,
 p-plus-s-out (no analyzer) combinations of polarization of the fundamental and second harmonic
(SH) waves were studied. The temperature of the LB films placed into the cryostat was varied in the
range from -30$^o$C up to 120$^o$C.

To characterize the general nonlinear optical properties of ultrathin ferroelectric LB films the
azimuthal anisotropy of the SHG intensity was studied. The top-left panel in Figure 1 shows the
dependence of the s-in, s-out SHG intensity on the azimuthal angle for a 15-monolayer-thick P(VDF-
TrFE) LB film. This dependence reveals an anisotropic component which possesses a distinct two-fold
symmetry riding on an isotropic background. The former is related to a coherent SHG and is
consistent with the presence of a predominant in-plane direction in the film symmetry. This
predominant direction can be attributed to a well ordered structure of oriented polymer chains in LB
monolayers which was observed previously for the same LB films by STM \cite{c3}. The latter is associated
with an incoherent SHG (the so-called hyper-Rayleigh scattering (HRS)) and is caused by random
inhomogeneity of spatial distribution of linear and non-linear optical parameters of LB films. Specific
features of HRS from thin inhomogeneous films has been discussed in details in recent studies \cite{10} and
in further presentation we will not distinguish between a coherent and incoherent SHG. Thus, in the
measurements of the SHG temperature dependence the azimuthal angle of the sample was set at the
maximum of an anisotropy of the SHG intensity. To collect the depolarized and diffuse SHG radiation,
the p-in, p-plus-s-out (no analyzer) combination was chosen and the SHG output was integrated over
an angular aperture of a detection system centered along the specular direction.

Figure 1 (main panel) shows the temperature dependence of the SHG intensity for a 60-
monolayer-thick P(VDF-TrFE) LB film. Two specific ranges in the SHG temperature dependence
could be marked out which are separated from each other at approximately 30$^o$C - 35$^o$C. As the
temperature is decreased in the first range, the dependence of the SHG intensity below 30$^o$C reveals
a pronounced peak at approximately 28$^o$C and a further monotonic increase of the nonlinear
response with the final saturation regime of the SHG signal below 10$^o$C. Heating above 35$^o$C (in the
second specific range) shows a slightly changing SHG signal with a gradual increase of the nonlinear
response until approximately 100$^o$C where saturation occurs. Cooling the sample in the second
specific range brings about a broad and strong hysteresis loop of the nonlinear response with a
maximum of the SHG intensity at approximately 80$^o$C. Similar features were observed in the
temperature dependence of the SHG intensity from a 15-monolayer-thick P(VDF-TrFE) LB film.

These specific features of the SHG temperature dependence for both temperature ranges can be
considered in terms of the ferroelectric-paraelectric phase transitions. It has been observed recently
that P(VDF-TrFE) LB films reveal two types of 2D ferroelectric-paraelectric phase transition \cite{c3,c7}.
The first one is a phase transition with a thickness independent Curie temperature T$_c^B$ in the vicinity of
80$^o$C and is related to a 3D ferroelectric state of a bulk P(VDF-TrFE) material \cite{c11}. This transition
possesses a wide temperature hysteresis of the static (low-frequency) dielectric constant, which shows
a significant increase on the cooling branch of the hysteresis loop (see the top-right panel in Fig. 1.).
The second first-order phase transition is observed at a Curier temperature T$_c^S$ of approximately 20$^o$C.
This transition is attributed to a ferroelectric ordering in the topmost surface layer of multilayered
LB structures of P(VDF-TrFE).

For a qualitative description of the temperature dependence of the SHG intensity in P(VDF-
TrFE) LB films, we consider a nonlinear-optical model of the ferroelectric LB structure shown in the
top panel of Fig.2. This model structure involves the non-linear optical sources located at the two
(air-LB and LB-substrate) interfaces and in the "bulk" of the LB film. The mid panel in Fig. 2 shows
schematically the model temperature dependences for the surface susceptibilities and their
interference. The Curie temperatures, T$_c^{S1}$ and T$_c^{S2}$, of ferroelectric-paraelectric phase transitions in
interfacial layers are slightly different, being in the vicinity of 20$^o$C, and the temperature dependent
surface dipole susceptibilities of second order, $\chi^{(2),S1}$(T) and $\chi^{(2),S2}$(T), have different absolute
values. In the paraelectric phase these surface susceptibilities originate from discontinuity of the
structure in the normal direction \cite{8} and are temperature independent. The phase shift between these
surface susceptibilities is supposed to be close to $\pi$ because of the opposite orientations of the
normal to these interfaces. The nonlinear optical properties of the bulk of ferroelectric LB films are
described by two, dipole and quadruple, bulk susceptibilities. The bottom panel in Fig. 2 shows a
schematic of the model temperature dependences for the bulk susceptibilities and their interference.
The dipole susceptibility,  $\chi^{(2),d}$(T), is temperature dependent and exists only below the bulk Curie
temperature, T$_c^B$, in the vicinity of 80$^o$C. The quadruple susceptibility, $\chi^{(2),Q}$, is temperature
independent, in the first approximation, and its phase is supposed to be shifted by p with respect to
the dipole bulk susceptibility,  $\chi^{(2),d}$(T). Recent SHG studies of LB films of large molecules such as
fullerene \cite{c10} do show a significant role of the nonlocal quadruple contribution to a quadratic
response of thin films.

The SHG intensity in our model is given by the following combination of quadratic nonlinear
polarizations of components of a planar structure, shown in Figure 2:

\begin{equation}
I_{2\omega}(T)\propto|$\overrightarrow{P}$_{2\omega}^{D,S1}(T)+$\overrightarrow{P}$_{2\omega}^{D,S2}(T)|^2+
         |$\overrightarrow{P}$_{2\omega}^{D,bulk}(T)+$\overrightarrow{P}$_{2\omega}^{Q,bulk}(T)|^2
         \text{,}  \label{eq1}
\end{equation}

\noindent where the interference of two surface, and two bulk contributions to the SH field is taken into
account. This model ignores the propagation factors, a multiple reflection interference which gives a
phase shift between interfering SH waves, and cross-interference of surface and bulk nonlinear
polarization terms, which should be considered in a more general model. The temperature
dependence of linear Fresnel factors is ignored in spite of the strong temperature dependence and
hysteresis of static dielectric constant $\varepsilon$(T) (see the top-right panel in Fig. 1). It is known that the
strong variations of static $\varepsilon$(T) do not result in the same strong variations of optical refractive index.
The latter does not allow to explain strong temperature dependence and hysteresis of the SHG
intensity.

An interpretation of the results of the SHG experiments is based on qualitatively comparing the
features of experimental SHG temperature dependences and the SHG intensity behavior in the
interference model summarized in Eq. 1. The interference of nonlinear polarizations of surface
ferroelectric layers and their temperature dependence under surface ferroelectric-paraelectric phase
transitions can give a peak in the SHG intensity in the vicinity of 20$^o$C and a gradual increase and
saturation in the SHG intensity for the low-temperature range (T $<$ 30$^o$C) (see mid panel in Fig. 2).

The interference of temperature independent quadruple polarization and temperature
dependent ferroelectric contribution in the bulk of the LB film, explains the heating branch of the
temperature dependence in Fig. 2 in the vicinity of T$_c^B$ $\approx$ 80$^o$C (see panel 3 in Fig. 2). The cooling
branch of the hysteresis loop in Fig. 1 shows a significant increase of the SHG intensity in the vicinity
of 80$^o$C. This hysteresis of the SHG intensity referred to an analogous hysteresis of the static $\varepsilon$(T)
can be explained by hysteresis of spontaneous polarization and corresponding SP-induced
contribution to the nonlinear-optical response.

In summary, an electrode-free method of optical SHG is used to study the 2D ferroelectric-
paraelectric phase transitions in LB films of P(VDF-TrFE). Temperature dependence of the SP-induced
 SHG intensity reveals two phase transitions: 2D ferroelectric phase transition in the interfacial
layers of the LB films and a thickness independent (almost 2D) ferroelectric phase transition in the
bulk of the LB films.

\newpage

\section*{Figure Captions}

\noindent Fig. 1. The main panel shows the temperature dependence of the SHG intensity for a 60-monolayer-thick
 LB film. Arrows show the direction of temperature variations. Filled and open symbols are for
the heating and cooling branches of the hysteresis loop, respectively. The top-left panel is the
azimuthal angular dependence of the SHG intensity from a 15-monolayer thick LB film. The top-right
panel is the temperature dependence of the static dielectric constant (after Ref. \cite{c3}).
\\

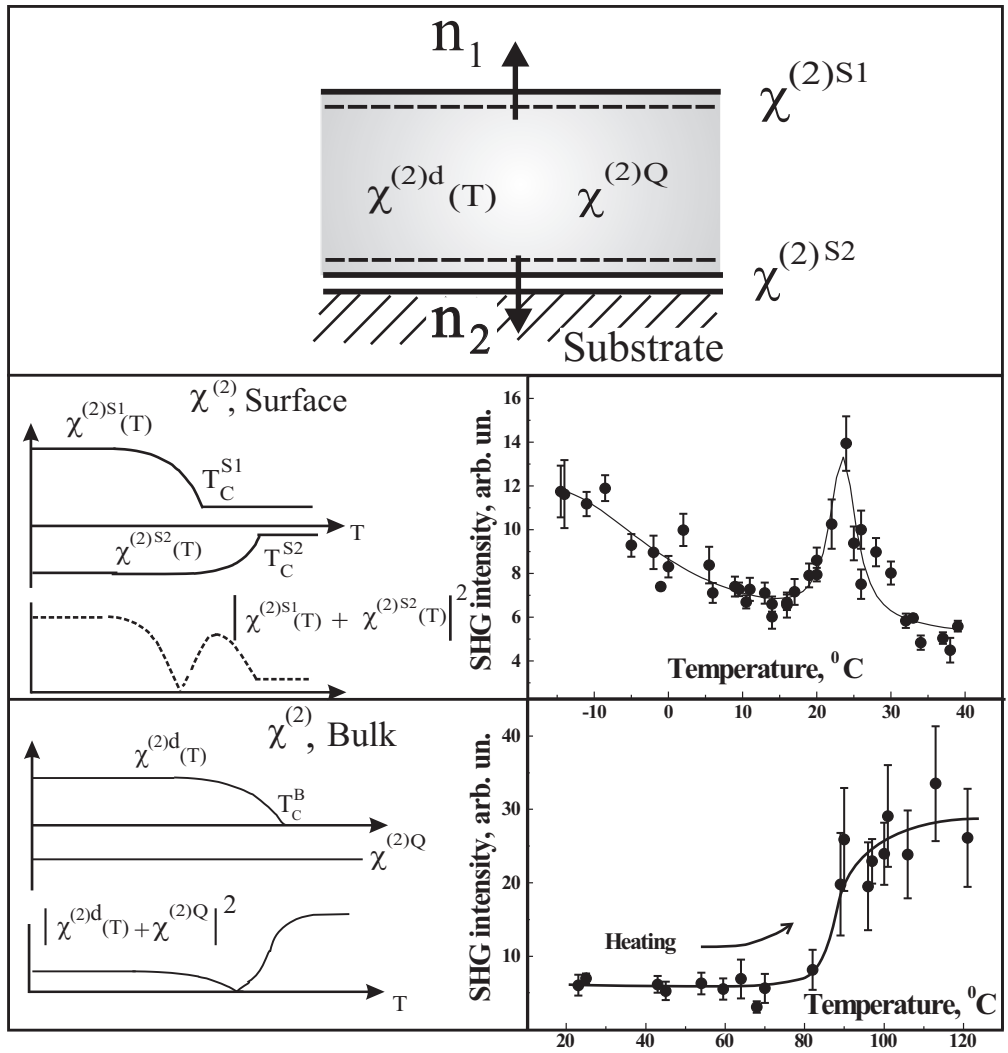
\noindent Fig. 2. The top panel shows a schematic of the nonlinear-optical model of ferroelectric LB film which
includes the air-LB film and LB film-substrate interfaces with dipole quadratic susceptibilities,
$\chi^{(2),S1}$(T) and $\chi^{(2),S2}$(T), with Curie temperatures T$_c^{S1}$, T$_c^{S2}$, and the bulk quadruple,
$\chi^{(2),Q}$, and
dipole, $\chi^{(2),d}$(T), susceptibilities below Curie temperature, T$_c^{B}$. The mid and bottom panels: the left
parts show, schematically, the model temperature dependences of the surface and bulk susceptibilities
and their interfering combinations for the model temperature dependence of the SHG intensity,
respectively; the right sides show the experimental temperature dependence of the SHG intensity in the
vicinity of surface phase transition for a 15-monolayer-thick LB film and of bulk phase transition for a
60-monolayer thick LB film, respectively.


\begin{thebibliography}{99}
\bibitem {c1} N. D. Mermin, H. Wagner, Phys. Rev. Lett. {\bf 17}, 1133 (1966).

\bibitem {c2} D. Weller, S.F. Alvarado, W. Gudat, K. Schroder, and M. Campagna,
Phys. Rev. Lett., {\bf 54}, 1555 (1985).

\bibitem {c3} A. V. Bune, V. M. Fridkin, S. Durcharme, L. M. Blinov, S. P. Palto,
A. V. Sorokin, S. G. Yudin, A. Zlatkin, Nature {\bf 391}, 874 (1998).

\bibitem {c4} G. G. Roberts, Ferroelectrics {\bf 91}, 21 (1989).

\bibitem {c5} G. J. Ashwell. P. D. Jackson, W. A. Crossland, Nature {\bf 368}, 438 (1994).

\bibitem {c6} M. Pomerantz, A. Aviram, Sol. St. Comm. {\bf 20}, 9 (1976).

\bibitem {c7} S. Ducharme, A. V. Bune, L. M. Blinov, V. M. Fridkin, S. P. Palto,
A. V. Sorokin, S. G. Yudin, Phys. Rev. {\bf B} {\bf 57}, 25 (1998).

\bibitem {c8} G. Lupke, Surf. Sci. Rep. 35, 75 (1999); J. F. McGilp, Phys. Stat. Sol. A 175, 153 (1999).

\bibitem {c9} O. A. Aktsipetrov, E. D. Mishina. Sov. Phys. Dokl. 29, 37 (1984).


\bibitem {c10} E. D. Mishina, T. V. Misuryaev, A. A. Nikulin, Th. Rasing, O. A. Aktsipetrov, J. Opt. Soc. Am. B
 {\bf 16}, 1692 (1999).

\bibitem {c11} J. F. Legrand, Ferroelectrics {\bf 91}, 303 (1989).
\end{thebibliography}
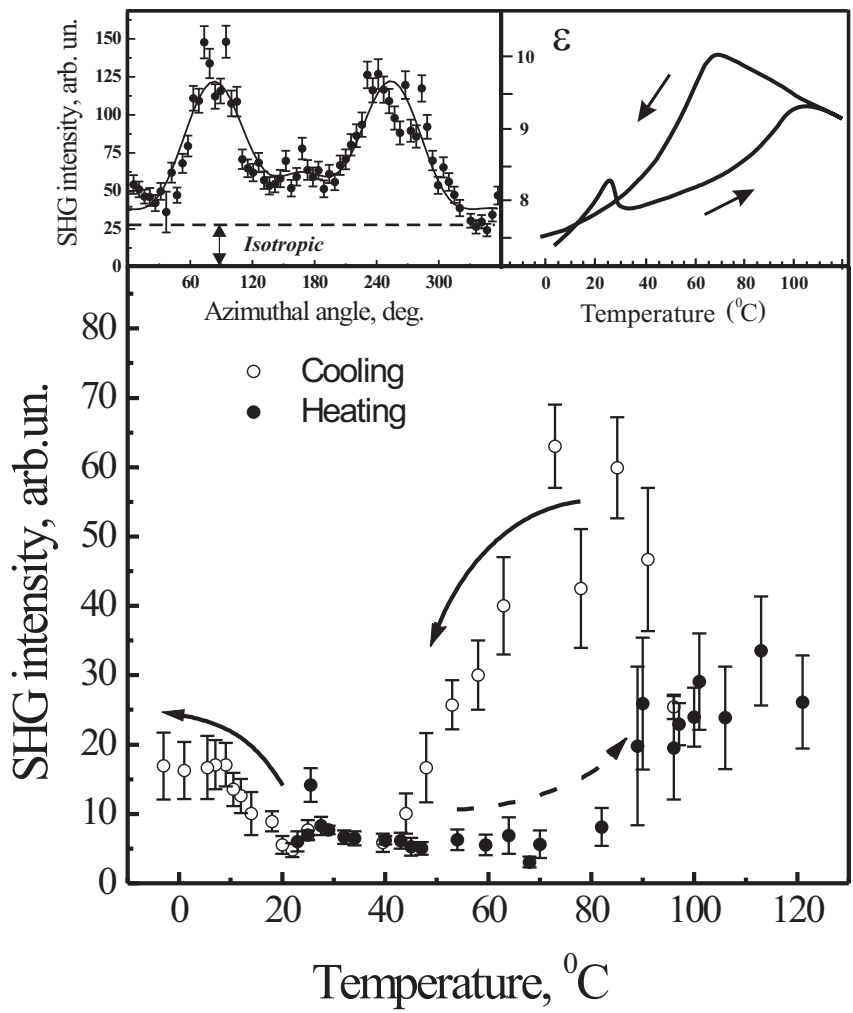
\end{document}